\documentclass[a4paper]{article}

\usepackage{INTERSPEECH2020}

\usepackage{multirow}
\usepackage{pifont}
\newcommand{\cmark}{\ding{51}}%
\newcommand{\xmark}{\ding{55}}%

\title{Streaming Transformer-based Acoustic Models Using Self-attention with Augmented Memory}
\name{
    Chunyang Wu, 
    Yongqiang Wang, 
    Yangyang Shi, 
    Ching-Feng Yeh,
    Frank Zhang
    \thanks{The authors would like thank for Christian Fuegen and Michael L. Seltzer for their support and discussion.}
}

\address{Facebook AI}

\begin{document}

\maketitle
\begin{abstract}
\vspace{0.5em}
  Transformer-based acoustic modeling has achieved great success for both hybrid and sequence-to-sequence speech recognition. However, it requires access to the full sequence, and the computational cost grows quadratically with respect to the input sequence length. These factors limit its adoption for streaming applications. In this work, we proposed a novel augmented memory self-attention, which attends on a short segment of the input sequence and a bank of memories. The memory bank stores the embedding information for all the processed segments. On the librispeech benchmark, our proposed method outperforms all the existing streamable transformer methods by a large margin and achieved over 15\% relative error reduction, compared with the widely used LC-BLSTM baseline. Our findings are also confirmed on some large internal datasets.
  
\end{abstract}

\section{Introduction}
Sequence modeling is an important problem in speech recognition. In both conventional hybrid \cite{bourlard2012connectionist} and end-to-end style (e.g., attention-based encoder-decoder \cite{bahdanau2016end, chiu2018state} or neural transducer\cite{he2019rnnt}) architectures, a neural encoder is used to extract a sequence of high-level embeddings from an input feature vector sequence. A feed-forward neural network extracts embeddings from a fixed window of local features \cite{hinton2012deep}. Recurrent neural networks (RNNs), especially the long short-term memory (LSTM) \cite{hochreiter1997long}, improve the embedding extraction by exploiting both long-term and short-term temporal patterns \cite{sak2014long}. Recently, attention (or self-attention if there is only one input sequence) has emerged as an alternative technique for sequence modeling \cite{vaswani2017attention}. Different from RNNs, attention connects arbitrary pairs of positions in the input sequences directly. To forward (or backward) signals between two positions that are $n$ steps away in the input, it only needs one step to traverse the network, compared with $O(n)$ steps in RNNs. Built on top of the attention operation, the transformer model \cite{vaswani2017attention} leverages multi-head attention and interleaves with feed-forward layers. It has achieved great success in both natural language processing \cite{devlin2018bert, radford2018improving} and speech applications \cite{karita2019comparative, wang2019transformer}.

However, two significant issues make transformer-based models impractical for online speech recognition applications. First, it requires access to the entire utterance before it can start producing output; second, the computational cost and memory usage grow quadratically with respect to the input sequence length if an infinite left context is used. There are a few methods that can partially solve these issues. First, {\it time-restricted self-attention}~\cite{povey2018time} can be used in which the computation of attention only uses the past input vectors and a limited length of future inputs (e.g.  \cite{zhang2020transformer, moritz2020streaming}). However, since the reception field is linearly growing for the number of transformer layers, it usually generates a significant latency; it does not address the issue of quadratically growing cost either. Second, {\it block processing} is used in \cite{dong2019self}, which chunks the input utterances into segments, and self-attention performs on each segment. In this way, the computation cost and memory usage don't grow quadratically. It is similar to context-sensitive-chunk BPTT in \cite{Chen2016chunk} and truncated BLSTM in \cite{mohamed2015deep}, which was successfully deployed to build online speech recognition system based on BLSTM models. However, since the transformer cannot attend beyond the current segment, it is observed that this method yields significant accuracy degradation \cite{Tsunoo2019, dai2019transformer}. Third, {\it recurrent connection}, in which embeddings from the previous segment are carried over to the current one, can be combined with the block processing. This approach is similar to the idea proposed in latency controlled BLSTM (LC-BLSTM) \cite{zhang2016highway}. An example of this approach is transformer-XL \cite{dai2019transformer}, in which it can model a very long dependency on text data for language modeling.  The work in  \cite{Tsunoo2019, tian2019synchronous} have explored similar ideas for acoustic modeling. 

Carrying over segment level information enables attention to access information beyond the current segment. A recurrent connection compresses the segment level information into a single memory slot. For a segment that is $k$ steps away, it takes $O(k)$ steps to retrieve the embedding extracted from that segment.  Inspired by the neural Turing machine\cite{graves2014neural}, we propose a novel {\it augmented memory} transformer, which accumulates the segment level information into a memory bank with multiple memory slots. Attention is then performed over the memory bank, together with the embeddings from the current segment. In this way, all the information, regardless of whether it is in the current segment or $k$ segments away, can be equally accessible. We applied this {\it augmented memory} transformer to hybrid speech recognition architecture and performed an in-depth comparison with other methods on a widely used LibriSpeech benchmark \cite{panayotov2015librispeech}. Experimental results demonstrate that the proposed augmented memory transformer outperforms all the other methods by a large margin. Using our proposed method, we show that with similar look-ahead sizes, augmented memory transformer improves over the widely used LC-BLSTM model by over 15\% relatively. Though we only evaluate the proposed method in a hybrid speech recognition scenario, it is equally applicable to end-to-end style architectures. 

The rest of this paper is organized as follows. In Section 2, we briefly review the self-attention and transformer-based acoustic model. We present the augmented memory transformer in Section 3. Section 4 demonstrates and analyzes the experimental results, followed by a summary in Section 5. 

\section{Transformer-based acoustic models}
\label{sec:am}
We first give a brief introduction of self-attention that is the core of the transformer-based model. Then we describe the architecture of the transformer-based acoustic model from \cite{wang2019transformer}. The model in this paper extends its model architecture for online streaming speech recognition.

\subsection{Self-attention}
Given an input embedding sequence $\boldsymbol{X} = (\boldsymbol{x}_1, ..., \boldsymbol{x}_T)$ where $\boldsymbol{x}_t \in \mathbb{R}^{D}$,
self-attention projects the input to \emph{query}, \emph{key} and \emph{value} space using $\mathbf{W}_{\rm q}$, $\mathbf{W}_{\rm k}$ and $\mathbf{W}_{\rm v}$, respectively,
\begin{align}
   \boldsymbol{Q}=\mathbf{W}_{\rm q}\boldsymbol{X},\ \ \  \boldsymbol{K}=\mathbf{W}_{\rm k}\boldsymbol{X},\ \ \ 
   \boldsymbol{V}=\mathbf{W}_{\rm v}\boldsymbol{X}
\end{align}
where $\mathbf{W}_{\rm q}, \mathbf{W}_{\rm k}, \mathbf{W}_{\rm v}$ are learnable parameters.
Self-attention uses dot-product to get the attention distribution over \emph{query} and \emph{key}, i.e., for position $t$ in \emph{query}, a distribution $\boldsymbol{\alpha}_{t}$ is obtained by: 
\begin{align}
    \alpha_{t\tau} = \frac{
    \exp(\beta \cdot\boldsymbol{Q}_t^{\sf T}\boldsymbol{K}_\tau )
    }{
    \sum_{\tau'} \exp(\beta \cdot\boldsymbol{Q}_t^{\sf T}\boldsymbol{K}_{\tau'} )}
\end{align}
where $\beta = \frac{1}{\sqrt{D}}$ is a scaling factor. Given $\boldsymbol{\alpha}_t$, the output embedding of self-attention is obtained via:
\begin{align}
    \boldsymbol{z}_t = \sum_{\tau} \mathrm{Dropout}(\alpha_{t\tau}) \cdot \boldsymbol{V}_\tau.
\end{align}
In \cite{vaswani2017attention}, multiple head attentions are introduced. Each of the attention head is applied individually on the input sequences. The output of each head is concatenated and linearly transformed into the final output.

\subsection{Transformer-based acoustic model}
The transformer-based acoustic model \cite{wang2019transformer} is a deep stack transformer layers on top of VGG blocks \cite{simonyan2014very}. Each transformer layer consists of a multi-head self-attention followed by a position-wise feed-forward layer. Rather than using Sinusoid positional embedding \cite{vaswani2017attention}, the transformer-based acoustic model \cite{wang2019transformer} uses VGG blocks to implicitly encode the relative positional information \cite{mohamed2019transformers}. The layer normalization \cite{lei2016layer}, the iterated loss \cite{Andros2019}, residual connections, and dropout is applied to train the deep stack transformer layers effectively. More model details can be found from \cite{wang2019transformer}.

\section{Augmented Memory Transformer}

The original transformer model generates the outputs according to the attention on the whole input sequence, which is not suitable for streaming speech recognition. The proposed augmented memory transformer addresses this issue by the combination of two mechanisms. 
First, similar to {\it block processing} \cite{dong2019self}, the whole utterance is segmented into segments padding with left context and right context. The size of each segment limits the computation and memory consumption in each transformer layer.
Second, to carry over information across segments, an {\it augmented memory bank} is used. Each slot in the {\it augmented memory bank} is the embedding representation of an observed segment.

\begin{figure}[t!]
    \includegraphics[width=3.2in]{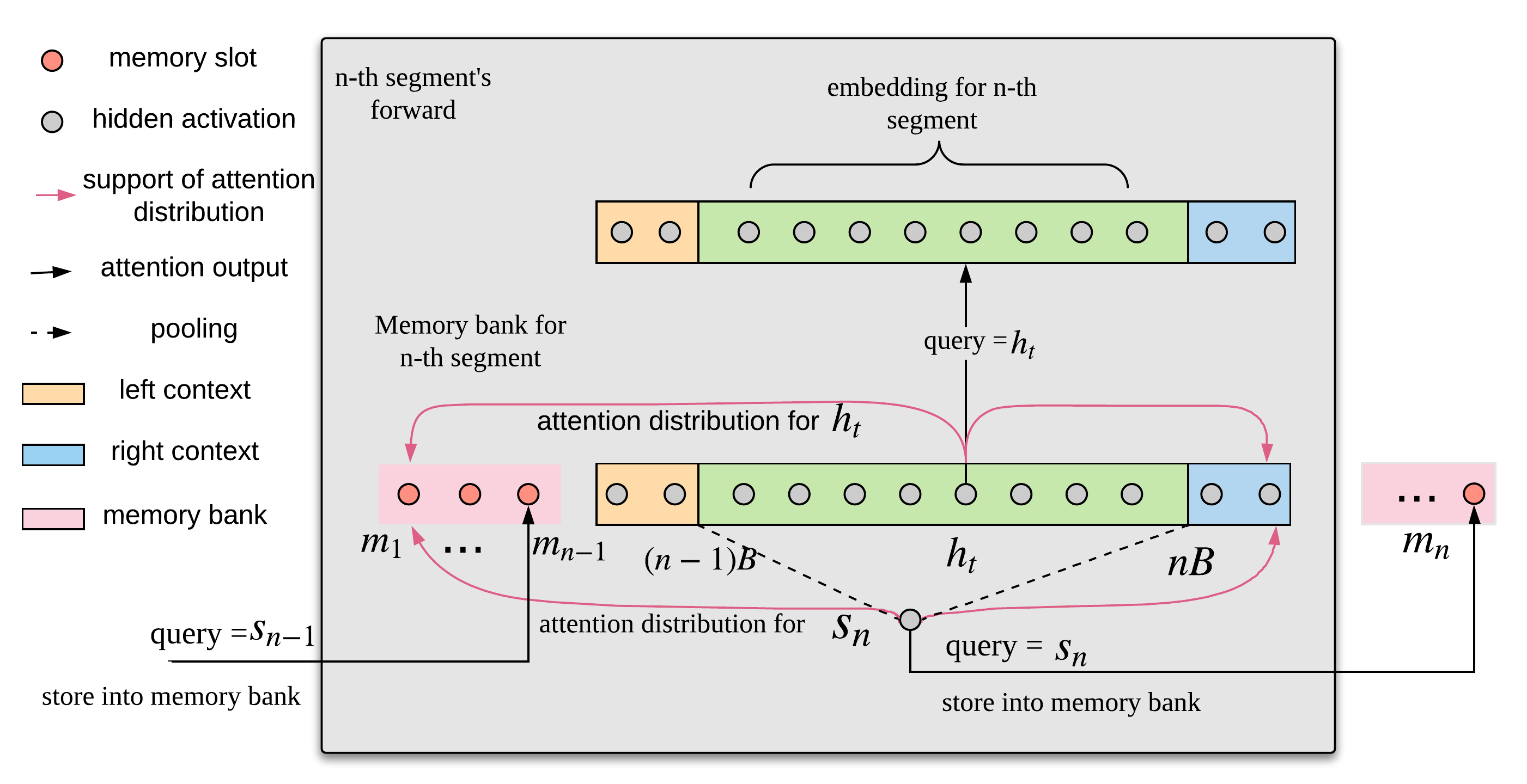}
    \label{fig:200_layer1}
    \caption{Illustration of one forward step for the augmented memory transformer on the $n$-th segment. }
    \label{fig:amtrf}
    \vspace{-1pt}
\end{figure}

Figure \ref{fig:amtrf} illustrates one forward step on the $n$-th segment using augmented memory transformer. 
An augmented memory bank (red) is introduced to the self-attention function.
The input sequence is first segmented into segments.
Each segment $\boldsymbol{C}_n=(\boldsymbol{x}_{nB+1}, ..., \boldsymbol{x}_{(n+1)B})$ contains $B$ input embedding vectors, where $B$ is referred to as the {\it segment length}.
The $n$-th segment is formed by patching the current segment with left context $\boldsymbol{L}_n$ (length $L$) and right context $\boldsymbol{R}_n$ (length $R$). An embedding vector $\boldsymbol{s}_n$,
referred to as the {\it summarization query} is then computed by pooling over $\boldsymbol{C}_n$.
Different pooling methods, e.g. average pooling, max pooling, and the linear combination, can be used.
This paper focuses on the average pooling. 
In the self-attention with augmented memory, the \emph{query} is the projection from the concatenation of current segment with context frames and the summarization query. The \emph{key} and the \emph{value} are the projections from the concatenation of the augmented memory bank and the current segment with context frames. They are formalized as
\begin{align}
     \boldsymbol{Q}&=\mathbf{W}_{\rm q}[\boldsymbol L_n, \boldsymbol{C}_n, \boldsymbol R_n , \boldsymbol{s}_n],\\
   \boldsymbol{K}&=\mathbf{W}_{\rm k}[\boldsymbol{M}_n, \boldsymbol{L}_n, \boldsymbol{C}_n, \boldsymbol{R}_n], \label{eq:amfk} \\
   \boldsymbol{V}&=\mathbf{W}_{\rm v}[\boldsymbol{M}_n, \boldsymbol{L}_n, \boldsymbol{C}_n, \boldsymbol{R}_n] \label{eq:amfv}
\end{align}
where $\boldsymbol{M}_n=(\boldsymbol{m}_1, ..., \boldsymbol{m}_{n-1})$ is the augmented memory bank. Note $\boldsymbol{Q}$ has $(L+C+R+1)$ column vectors and $\boldsymbol{q}_{-1}$ is the projection from $\boldsymbol{s}_n$. The attention output for $\boldsymbol{q}_{-1}$ is stored into augmented memory bank as $\boldsymbol{m}_{n}$ for future forward steps, i.e., 
\begin{align}
    \boldsymbol{m}_n = \sum_{\tau} \mathrm{Dropout}(\alpha_{(-1)\tau}) \cdot \boldsymbol{V}_\tau
\end{align}
where $\alpha_{(-1)\tau}$ is the attention weight for $\boldsymbol{q}_{-1}$. The attention output from $(\boldsymbol{q}_{1}, ..., \boldsymbol{q}_{L+C+R})$ is feed to the next layer, except for the last transformer layer, only the center $B$ vectors are used as the transformer network's output.
The output for the whole utterance is the concatenation of outputs from all the segments. 

The proposed method is different to existing models in a variety of aspects.
Transformer-XL~\cite{dai2019transformer} incorporates 
 history information only from previous segment $\boldsymbol{C}_{n-1}$ via 
\begin{equation}
\begin{split}
   \boldsymbol{Q}&=\mathbf{W}_{\rm q}\boldsymbol{C}_n,\\ \boldsymbol{K}&=\mathbf{W}_{\rm k}[\boldsymbol{C}_{n-1},\boldsymbol{C}_n],
   \boldsymbol{V}=\mathbf{W}_{\rm v}[\boldsymbol{C}_{n-1},\boldsymbol{C}_n].
\end{split}
\end{equation}
Also note that, in transformer-XL, $\boldsymbol C_{n-1}$ is from the lower layer. This makes the upper layers have an increasing long reception field. Our proposed augmented memory transformer explicitly holds the information from all the previous segments (Eq. \ref{eq:amfk} and \ref{eq:amfv}) and all the layers have the same reception field. Using a bank of memories to represent past segments is also explored in \cite{rae2019compressive}, primarily in language modeling tasks.
In \cite{povey2018time}, the {\it time-restricted transformer} restricts the attention to a context window in each transformer layer. This means the look-ahead length is linearly growing by the number of transformer layers. Our proposed method has a fixed look-ahead window, thus enable us to use many transformer layers without increasing look-ahead window size.

\section{Experiments}
The proposed model was evaluated on the LibriSpeech ASR task, and two of our internal video ASR tasks, German and Russian.
Neural network models were trained and evaluated using an in-house extension of the PyTorch-based \emph{fairseq}~\cite{ott2019fairseq} toolkit.
In terms of latency,
this paper focuses on the algorithmic latency,
i.e. the size of look-ahead window.
Different models were compared with similar look-ahead windows, for a fair comparison.

\subsection{LibriSpeech}
We first performed experiments on the LibriSpeech task \cite{panayotov2015librispeech}. 
This dataset contains about 960 hours of read speech data for training, and 4 development and test sets (\texttt{\{dev, test\} - \{clean,other\}}) for evaluation, where \texttt{other} sets are more acoustically challenging.
The standard 4-gram language model (LM) with a 200K vocabulary was used for all first-pass decoding. 
In all experiments,
80 dimensional log Mel-filter bank features with a 10ms frame-shift were used as input features.
The context- and position-dependent graphemes, i.e. \textit{chenones}~\cite{le2019senones}, were used as output labels.

\subsubsection{Experiment Setups}
\label{sec:libri_setup}

A GMM-HMM system was first trained following the standard Kaldi~\cite{Povey_ASRU2011} Librispeech recipe.
To speed up the training of neural networks,
the training data were segmented into utterances that were up to 10 seconds\footnote{The training-data segmentation was obtained from the alignments of an initial LC-BLSTM model.
According to our studies,
shorter segments in training can both improve the training throughput and decoding performance.};
speed perturbation \cite{ko2015audio} and \emph{SpecAugment} \cite{park2019specaugment} were performed on the training data.
In evaluation, no segmentation was performed on the test data.
This paper focuses on cross-entropy (CE) training for neural network models.
The proposed augmented memory transformer (AMTrf) was compared with streamable baselines including LC-BLSTM~\cite{zhang2016highway}, 
transformer-XL (Trf-XL)~\cite{dai2019transformer} 
and time-restricted transformer (TRTrf)~\cite{povey2018time}. Also, the non-streamable original transformer (Trf) was included to indicate potential performance lower-bound.

We started with investigating models of a small configuration with approximately 40M parameters.
The LC-BLSTM baseline consists of 5 layers with 400 nodes in each layer each direction. Mixed frame rate \cite{peddinti2017low}, i.e. the output frames of the first layer are sub-sampled by factor of 2 before propagated to the the second layer, is used. The look-ahead window is set to 0.4 second, i.e. 40 frames; the chunk size in LC-BLSTM is 1.5 seconds.
For transformers, the topology is 12 transformer layers with 512 input embedding dimensions, 8 multi-heads, and 2048 feed-forward network (FFN) dimensions in each layer.
Following \cite{wang2019transformer}, two VGG blocks \cite{simonyan2014very} are introduced as lower layers before the stacked transformer layers\footnote{As studied in \cite{wang2019transformer}, the VGG blocks are a best practice of input positional embedding for transformers.
In experiments using VGG blocks on LC-BLSTM, insignificant gains obtained.}.
Each VGG block consists of two consecutive 3-by-3 convolution layers, a ReLu activation function, and a max-pooling layer; the first VGG block includes 32 channels, and the second VGG block has 64; 2-by-2 max-pooling is used in each block with stride 2 in the first and stride 1 in the second. The VGG blocks generate a 2560-D feature sequence at a 20ms frame rate.
In training,
the Adam optimizer \cite{kingma2014adam} was used for all the models.
Dropout was used: 0.5 for LC-BLSTMs and 0.1 for transformers.
The LC-BLSTM baseline was optimized for at most 30 epochs on 16 Nvidia V100 GPUs.
The learning rate was initially $10^{-4}$ and reduced 50\% after each epoch with accuracy degradation on the cross-validation data.
Transformer models were optimized using a tri-stage learning-rate strategy: 
8K updates with a learning rate increased linearly from $10^{-5}$ to a holding learning rate $3\times10^{-4}$, 100K updates with the holding learning rate, and further updates with the learning rate decreased exponentially.
32 GPUs were used in training one transformer model.
Transformer models were updated up to 70 epochs.

The large configuration, i.e. approximately 80M parameters, was then investigated.
The large LC-BLSTM baseline consists of 5 layers with 800 nodes in each layer each direction.
The large transformer consists of 24 layers. The layer setting is identical to that of the small configuration; also, the same VGG blocks are used.
The training schedule of LC-BLSTM and transformers followed a similar fashion as that in the small configuration.
For large transformers,
to alleviate the gradient vanishing issue, iterated loss~\cite{Andros2019} is applied. 
The outputs of the 6/12/18-th transformer layers are non-linearly transformed (projected to a 256-dimensional space with an linear transformation followed by a ReLU activation function), and auxiliary CE losses are calculated separately. These additional losses are interpolated with the original loss with an 0.3 weight.

In evaluation, 
a fully-optimized, static 4-gram decoding graph built by Kaldi was used. 
The results on test sets were obtained using the best epoch on the development set\footnote{A model that averaged the last 10 epochs was included as a model candidate.}.
Following \cite{luscher2019rwth},
the best checkpoints for \texttt{test-clean} and \texttt{test-other}
are selected the respective development sets.

\subsubsection{Segment and Context Length}

We investigate the effect of segment length and context size first.
A key issue on the proposed model is how to compromise between latency and accuracy.

\begin{table}[htb]
    \centering
    \caption{Effect of segment, left and right context length on \emph{LibriSpeech}. Length is measured by number of frames, where frames are shifted in a 10ms frame rate. }
    \begin{tabular}{|ccc||cc|}
    \hline
     Left & Segment & Right & test-clean & test-other \\
    \hline
    \hline
    0 & 64 & 0 & 10.7 & 13.9 \\
    0 & 96 & 0 & 9.8 & 13.0 \\
    0 & 128 & 0 & 7.7 & 10.4 \\
    \hline
    16 & 128 & 0  & 5.2 & 9.5 \\
    32 & 128 & 0 & 3.6 & 8.5 \\
    64 & 128 & 0 & 3.5 & 8.5 \\
    \hline
    0 & 128 & 16 & 5.5 & 9.3 \\
    0 & 128 & 32 & 3.8 & 8.1 \\
    \hline
    32 & 128 & 32 & 3.3 & 8.0 \\
    64 & 128 & 32 & 3.3 & 7.6 \\
    \hline
    --& $\infty$ & -- & 3.1 & 7.1 \\
    \hline
    \end{tabular}
    \label{tab:context_libri}
\end{table}
The decoding performance is reported in Table~\ref{tab:context_libri}.
The first block shows the results without context.
By increasing the segment length from 64 to 128 frames, the word error rate (WER) decreased.
Next, various context settings were investigated with the segment length fixed to 128 frames.
The second and third blocks illustrate
the effect of left and right contexts, respectively.
Either left or right contexts contributed to alleviating the boundary effect.
A more extended context was shown to improve the recognition accuracy.
Finally, the effect of using both contexts was shown in the fourth block.
The left and right contexts showed some level of complementarity; thus, the performance further improved.
The $\infty$ system refers to an transformer-based acoustic model presented in \cite{wang2019transformer}, indicating the performance lower-bound.

The setting of 128 segment length, 64 left, and 32 right contexts were investigated in the following experiments.
It yields a look-ahead window of 32 frames, i.e. 0.32 seconds, which is comparable to that of the LC-BLSTM baseline.

\subsubsection{Limited Memory}

The second set of experiments investigated
the effect of limited memory size.
Instead of the complete observation of augmented memory bank,
models in this section were trained and tested by observing a fixed number of the most recent memory vectors. Note, when memory size equals to 1, our methods becomes almost the same as the encoder used in \cite{Tsunoo2019}. 
These experiments were performed to investigate how much long-term history contributed to the final performance.
\vspace{-0.5em}
\begin{table}[htb]
    \centering
    \caption{Effect of limited memory size on \emph{LibriSpeech}.}
    \begin{tabular}{|c||cc|}
    \hline
    MemSize & test-clean & test-other \\ 
    \hline
    \hline
    0 & 3.2 & 8.1 \\
    1 & 3.3 & 8.0 \\
    3 & 3.2 & 7.9 \\
    5 & 3.3 & 7.9 \\
    $\infty$ & 3.3 & 7.6 \\
    \hline
    \end{tabular}
    \label{tab:mem_libri}
\end{table}
Table~\ref{tab:mem_libri} reports the results using different memory sizes.
On the noisy set \texttt{test-other},
the performance was consistently improved
from no memory (0) to unlimited memory ($\infty$)\footnote{The longest utterance in the LibriSpeech test sets is about 35 seconds. Thus, the $\infty$ system used maximum 28 memory slots.}.
However on the clean data \texttt{test-clean},
little improvement was obtained.
This observation indicates that the global information in long-term memory
can alleviate more challenging acoustic conditions.

\subsubsection{Comparison with Other Streamable Models}

Table~\ref{tab:cmp_libri} compares the WERs of different models.
For a fair comparison on latency,
corresponding models of similar look-ahead window are compared.
The first block compares
models with about 40M parameters.
The transformer-XL baseline used a segment length of 128, which is identical to that of the proposed model.
The "+look-ahead" reports the extension of transformer-XL with right context\footnote{There is no context in the original design of transformer-XL. We applied a similar idea of right context (32 frames) on transformer-XL as the proposed model. Thus, both model has a look-ahead window of 0.32 second.}.
The TRTrf baseline used a context of 3 in each layer, resulting a look-ahead window of 0.72 second.
The proposed augmented memory transformer outperformed all the streamable baselines.
\begin{table}[htb]
    \centering
    \caption{Performance of different models on \emph{LibriSpeech}. ``Str'' stands for streamable, specifying if a model is a streamable one.}
    \begin{tabular}{|c|c|l||cc|}
    \hline
    \#Param & Str &  Model   & test-clean & test-other \\
    \hline
    \hline
    \multirow{6}{*}{$\simeq$40M} & \multirow{5}{*}{\cmark} & LC-BLSTM  & 3.8 & 9.9 \\
    & & Trf-XL   & 4.2 & 10.7 \\
    & & \ \ \ \ +look-ahead & 3.9 & 10.1 \\
    & & TRTrf & 4.1 & 9.0 \\
    & & AMTrf  & \textbf{3.3} & \textbf{7.6} \\
    \cline{2-5}
    & \xmark & Trf & 3.1 & 7.1 \\
    \hline
    \hline
    \multirow{6}{*}{$\simeq$80M}& \multirow{5}{*}{\cmark} & LC-BLSTM  & 3.3 & 8.2 \\
    & & Trf-XL& 3.5 & 8.3 \\
    & & \ \ \ \ + look-ahead& 3.2 & 7.7 \\
    & & AMTrf & 3.1 & 7.1 \\
    & & \ \ \ \ +WAS  & \textbf{2.8} & \textbf{6.7} \\
    \cline{2-5}
    & \xmark & Trf & 2.6 & 5.6 \\
    \hline
    \end{tabular}
    \label{tab:cmp_libri}
\end{table}
Larger models with about 80M parameters are compared in the second block.
The augmented memory transformer shows consistent gains as the small-size one.
For further improvement,
the weak-attention suppression (WAS) \cite{yang2020weak} was applied on top of the proposed model, denoted by "+WAS".
Compared with the LC-BLSTM baseline,
the augmented memory transformer (with WAS) achieved 15\%-18\% relative error reduction on the two test sets.
At the time of writing, this is the best number that we acknowledge on LibriSpeech for streamable models.

\subsection{Video ASR}
To evaluate the model in more challenging acoustic conditions, our in-house Russian and German video ASR datasets were used.
The videos in this dataset are originally shared publicly by users; only the audio part of those videos are used in our experiments. These data are completely de-identified; both transcribers and researchers do not have access to any user-identifiable information.
For the Russian task,
the training data consisted of 1.8K hours from 100K video clips.
14.6 hours of audio (790 video clips) were used as validation data.
Two test sets were used in evaluation:
the 11-hour \texttt{clean} (466 videos),
and 24-hour \texttt{noisy} (1.3K videos) sets.
For the German task,
the training data consisted of 3K hour audios (135K videos).
The validation data was 14.5 hours (632 videos).
The test data were 
the 25-hour \texttt{clean} (989 videos) and
 24-hour \texttt{noisy} (1K videos) sets.

\vspace{-0.5em}
\begin{table}[htb]
    \centering
    \caption{Experiment results on our internal \emph{video ASR} tasks.}
    \begin{tabular}{|c|l||cc|}
    \hline
    Language & Model & clean & noisy  \\
    \hline
    \hline
    \multirow{3}{*}{Russian} & LC-BLSTM & 19.8 & 24.4 \\
    & AMTrf & \textbf{18.0} & \textbf{23.3} \\
    \cline{2-4}
    & Trf &  16.6 & 21.1  \\
    \hline
    \hline
    \multirow{3}{*}{German} & LC-BLSTM & 19.6 & 19.5 \\
    & AMTrf & \textbf{17.4} & \textbf{17.1}  \\
    \cline{2-4}
    & Trf & 16.2 & 15.6  \\
    \hline
    \end{tabular}
    \label{tab:video_asr}
\end{table}

\vspace{-0.5em}
The large network configuration, i.e. 80M-parameter models, was examined.
The training of all the models was performed in a similar fashion as presented in Section~\ref{sec:libri_setup} (large configuration).
Table~\ref{tab:video_asr} summarizes the decoding results.
On both languages,
the proposed model consistently outperformed the LC-BLSTM baseline by 9-11\% on clean test sets and 5-12\% on noisy test sets.  There are still some accuracy gaps compared with the transformer which has the access to the whole utterance. 

\vspace{-0.5em}
\section{Conclusions}
In this work, we proposed the augmented memory transformer for streaming transformer-based models for speech recognition. 
It processes sequence data incrementally using short segments and an augmented memory, thus has the potential for latency-constrained tasks.
On LibriSpeech, the proposed model outperformed LC-BLSTM and all the existing streamable transformer baselines. Initial study on more challenging Russian and German video datasets also illustrated similar conclusions.

In this paper, the latency was measured in an algorithmic way, i.e. look-ahead window size;
we will investigate the real latency and measure the throughput of this model.
The proposed method can be also applied to transformer transducer \cite{zhang2020transformer, yeh2019transformer} or transformer-based sequence-to-sequence models (e.g.  \cite{mohamed2019transformers, karita2019comparative}).

\bibliographystyle{IEEEtran}

\bibliography{mybib}


\end{document}